\overfullrule=0pt
\baselineskip= 18pt plus 3pt

\input epsf


\def\picture #1 by #2 (#3){
  \vbox to #2{
    \hrule width #1 height 0pt depth 0pt
    \vfill
    \special{picture #3} 
    }
  }
\def\scaledpicture #1 by #2 (#3 scaled #4){{
  \dimen0=#1 \dimen1=#2
  \divide\dimen0 by 1000 \multiply\dimen0 by #4
  \divide\dimen1 by 1000 \multiply\dimen1 by #4
  \picture \dimen0 by \dimen1 (#3 scaled #4)}
  }

\def\FigureA{\scaledpicture 9.03 in by 2.96 in (FigureA scaled 800)}
\def\FigureB{\scaledpicture 6.90 in by 7.19 in (FigureB scaled 800)}
\def\FigureCA{\scaledpicture 7.40 in by 9.89 in (FigureCA scaled 800)}
\def\FigureCB{\scaledpicture 6.94 in by 9.71 in (FigureCB scaled 900)}
\def\FigureCC{\scaledpicture 3.90 in by 9.71 in (FigureCC scaled 800)}
\def\FigureD{\scaledpicture 6.50 in by 5.68 in (FigureD scaled 800)}
\def\FigureE{\scaledpicture 6.93 in by 8.82 in (FigureE scaled 850)}
\def\FigureF{\scaledpicture 5.36 in by 7.54 in (FigureF scaled 950)}
\def\FigureG{\scaledpicture 8.19 in by 7.04 in (FigureG scaled 800)}

\def\TableA{\scaledpicture 7.86 in by 9.57 in (TableA scaled 850)}
\def\TableB{\scaledpicture 7.31 in by 3.22 in (TableB scaled 800)}

{\bf
\centerline{Automated and accurate protein structure description:}
\centerline{ Distribution of Ideal Secondary Structural Units in
Natural Proteins} 
\centerline{ }
}

\vskip 2cm \centerline{Sushilee Raganathan,$^a$ Dmitry Izotov,$^a$ Elfi Kraka,$^{a}$ Êand 
Dieter Cremer$^{a,b,*}$} 

\medskip \centerline {$^a$Department of Chemistry,
University of the Pacific}
\centerline {3601 Pacific Ave., ÊStockton, CA 95211-0110}

\medskip \centerline {$^b$Department of Physics,
University of the Pacific}
\centerline {3601 Pacific Ave., ÊStockton, CA 95211-0110}

\vskip 3cm

\parskip = 6pt plus 1pt

\noindent {\bf Abstract:} A new method for the {\it A}utomated {\it P}rotein {\it S}tructure
{\it A}nalysis (APSA) is derived, which simplifies the protein backbone to a smooth curve in
3-dimensional space. For the purpose of obtaining this smooth line  each amino acid is represented by its C$_{\alpha}$ atom, which serves as suitable anchor point for a
cubic spline fit. The backbone line is  characterized by arc length $s$, curvature $\kappa(s)$, and torsion
$\tau(s)$. The $\kappa(s)$ and $\tau(s)$ diagrams of the protein backbone suppress, because of the level of coarse graining
applied, details of the bond framework of the backbone, however reveal accurately all  secondary structure features  of a
protein. Advantages of  APSA are its quantitative representation and analysis of 3-dimensional structure in form of 2-dimensional curvature and torsion patterns, its
easy visualization of complicated conformational features, and its general applicability. Typical differences between 3$_{10}$-,
$\alpha$-, 
$\pi$-helices, and
$\beta$-strands    are quantified with the help of the  $\kappa(s)$ and $\tau(s)$ diagrams. For a test set of 20 proteins, 63 \% of all helical residues and
48.5 \% of all extended residues are identified to be in ideal conformational environments with the help of APSA. APSA is compared with other  methods for protein
structure analysis and its  applicability to higher levels of protein structure is discussed.

\bigskip
\noindent {\it key words:} protein structure, protein backbone, curvature, torsion, secondary structure, helices, $\beta$-sheets

\parskip = 6pt plus 1pt
\vfill
\eject

{\bf 1. Introduction}
\smallskip

Protein structure is organized into primary, secondary, tertiary, and quaternary levels expressing in this way its
enormous variety and complexity. [1] There have been numerous attempts of
simplifying measured structures for the purpose of finding unique conformational patterns.  For example, Venkatachalam  [2]
presented a local description of a protein molecule based merely on the polypeptide chain backbone. Furthermore, he showed that
 just two of the three conformational angles of the backbone have to be specified  for
a particular amino acid residue because conjugative effects keep the peptide unit planar. [2] These simplifications do not hinder
the valid description of a protein, but confirm earlier predictions [1] of its periodical structure  in agreement with crystallographic data. [3]  A different
approach employed by Kabsch and Sander [4] described secondary structures in terms of the shape and organization of hydrogen-bonded rings found along the backbone. They
were able to identify helices and beta sheets quickly and precisely. A higher level of organization that involved pairs of secondary structures called  the supersecondary
structures was identified by Rao and Rossman [5] by comparison of  protein structures.  These and other methods were applied for protein structure description, always
with the objective of developing appropriate tools for protein structure prediction. From an elementary point of view one can distinguish between three different
approaches to fulfill these objectives. One can base protein structure analysis and prediction exclusively on conformational (geometrical) features. Alternatively, 
one can correlate  structural features with   amino acid properties such as  H-bonding ability, hydrophilicity, hydrophobicity, polarity, etc. [1] and use the latter
properties for structure classification. Finally, a combination of conformational (geometrical) and physiochemical amino acid properties  can be used for the purpose of
structure analysis and prediction.

Based on various such descriptions, secondary structure elements like helices and $\beta$ sheets, supersecondary structures like
hairpins and corners, larger supersecondary motifs like the beta barrel and folds of domains in globular proteins have been
extensively classified. [1] Classification of proteins deposited in the Protein Data Bank (PDB) [6] can be found in databases such as SCOP [7], CATH [8], DALI [9],  etc.,
which combine  automated and manual sorting of domains. Such domain classifications have  been compared and analyzed [10] and it has been shown that they are frequently
conflicting with regard to domain definition and assignment of domain boundaries. Conflicting descriptions can also be obtained when using secondary structure assignment
programs such as DSSP [4], STRIDE [11], DEFINE [12], and KAKSI [13] (for a comparison, see Ref. [13]). It requires detailed and individual analysis of
structures at specific localized sites to classify some of the loop regions and show them to be made up of simpler units [14,15].
 More recently, secondary structure has been analyzed
on the basis of neural networks. [16] In general, modern secondary structure assignment and prediction can refer to a multitude of
strategies  based on H-bonding [4], backbone conformation [2], multiple sequence alignments [17], energy based evaluations
[18], etc.

It is generally accepted that one can recognize protein structure from the form of its backbone. [1,2] Once the
backbone structure is understood, one can complete the backbone by adding side chains using available procedures. [19] Despite
such  accomplishments, the need for improved  analysis of backbone structure is  mentioned [20] in connection with
protein structure predictions and when elucidating protein functionality.  [1] Hence an automated systematic description of the
backbone structure of a protein, is still, after many decades of protein studies, a needed tool to relate protein
primary structure to protein functionality. The state of the art in protein structure description is often judged in view of its
value for protein structure prediction where the efficiency of automation  plays a major role. [21]

Any useful description of backbone or total protein structure has to consider the different levels of structural
description, which lead from the secondary elements through the supersecondary units and motifs to the tertiary level with folds of domains. [1] This hierarchy of
structural descriptors implies that for any given point of a protein backbone an increasingly larger environment has to be taken into account to step up from a purely
local to a  global description. Clearly, any complete description of the protein backbone must include all levels of the structural hierarchy. One could
assume that after 50 years of research in this field at least the secondary level of protein structure is well-understood; however this is questionable given that
even  helices are still under evaluation. [22,23]

\bigskip


\epsfbox {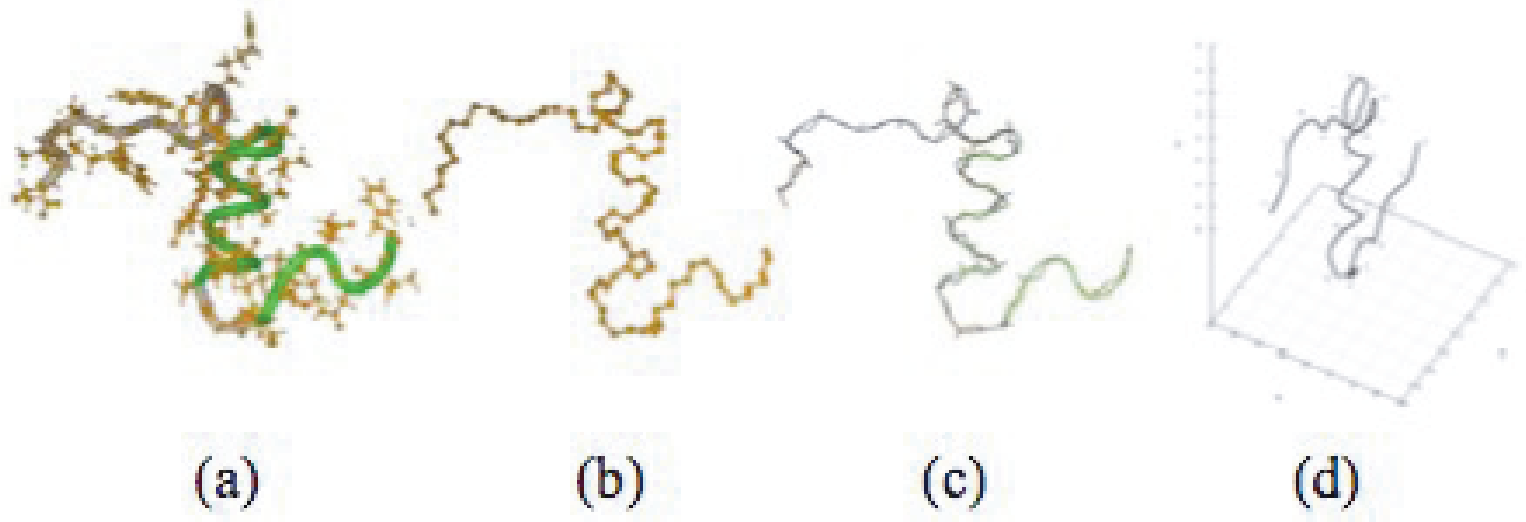}

\noindent {\bf Figure 1}. Simplification of the protein structure to a smooth line in 3D-space in 3 steps. (a) Perspective drawing of the
structure of insulin B chain taken from PDB file 1JCO.  [6] (b) After deleting all side chains the backbone of insulin B chain becomes visible.
The C$_\alpha$ atoms of all residues are indicated by dots. (c)  The positions of the C$_\alpha$ atoms are connected by a cubic spline fit. (d)
The backbone is represented as a smooth line in 3D-space.

\bigskip

In this work, we will approach the task of describing protein structure in
the spirit of early structural simplifications. We will delete all side chains and focus just on the protein backbone (Figures
1a to 1d). The orientation of the backbone in 3-dimensional (3D) space can be determined by a multitude of residue dihedral angles
$\phi$ and $\psi$.  However, such an approach  leads to a local and discrete description of the backbone that, only
with difficulties, can be applied for the description of tertiary structure. For the purpose of developing a simpler approach to
protein structure description we introduce two strategies: 

a) We sacrifice a fine-grained atom-by-atom description of the backbone leading to excessive detail to a coarse-grained one. 
The latter can be done in various steps with the purpose of including increasingly more non-local structural features into the
protein description. At the level used in this work, each residue is represented by just one anchor point 
(Figure 1c).  At higher levels of coarse graining, one would have
to present supersecondary or tertiary structural units by new anchor points to add more non-local features to the protein
description. In this way, all features from secondary to tertiary structure become accessible to a purely conformational (geometric) approach of
structure description.  

b) The second feature of the current approach is based on a continuous rather than discrete representation of the
protein backbone. We convert the latter into a smooth line oriented in 3D space by the anchor points of step (a).
Along this line, we use three mathematical parameters to determine length and orientation of the backbone line, namely the arc
length $s$, the scalar curvature $\kappa$, and the scalar torsion $\tau$. By expressing $\kappa$ and $\tau$
as a function of the arc length $s$ we obtain a continuous representation of the protein backbone (see Figure 1d), which adopts
characteristic patterns for given structural features of a protein. The advantages of such a strategy lie in its ease of
automation in form of a computer program, its general applicability, and the mathematical accuracy of structure description, as  will be shown in this work.

A detailed and systematic presentation of the approach sketched above and  results obtained with it are given in the following sections. In Section 2, mathematical
procedures and the data set used in this work are described. Results and discussion will be presented in Section 3. Section 4
is devoted to a comparison of the current method with other methods of protein structure description and prediction. Finally,
we will summarize the most important conclusions of this work and provide an outlook on the next steps of this work.

\bigskip
\noindent {\bf 2. Computational methods and procedures}
\smallskip

As described in the introduction, the protein backbone can be presented as a smooth regularly parameterized  space-curve. Any
sufficiently smooth curve in 3D space is completely characterized by three parameters including its arc length
$s$,  scalar curvature $\kappa$, and scalar torsion $\tau$ where $\kappa$ and $\tau$ are expressed as a function of $s$. [24] In
Figure 2, tangent vector {\bf t}, normal vector {\bf n} (in applications also called curvature vector), and binormal vector {\bf b} of a Frenet frame at point P$_1$ of a
left-handed helix curve {\bf r}(s)  are shown. If the Frenet frame moves along the curve to point P$_2$, {\bf t}, {\bf n}, and {\bf b} readjust their orientation where
the rotation of the Frenet frame around vectors  {\bf b} and {\bf t} is given by  curvature and torsion. The formulas for the latter quantities
expressed as a function of the arc length {\it s} are: [24]  
$$
\kappa(s) = |{\bf r}''(s)|, \eqno(1)
$$
$$
\tau(s) = {{\langle {\bf r}'(s),{\bf r}''(s),{\bf r}'''(s)\rangle}
\over {|{\bf r}''(s)|^2}}, \eqno(2)
$$
where 
$|\cdot|$ is the norm, and $\langle\cdot\rangle$ is the
triple product.

\midinsert


\centerline{\epsfbox {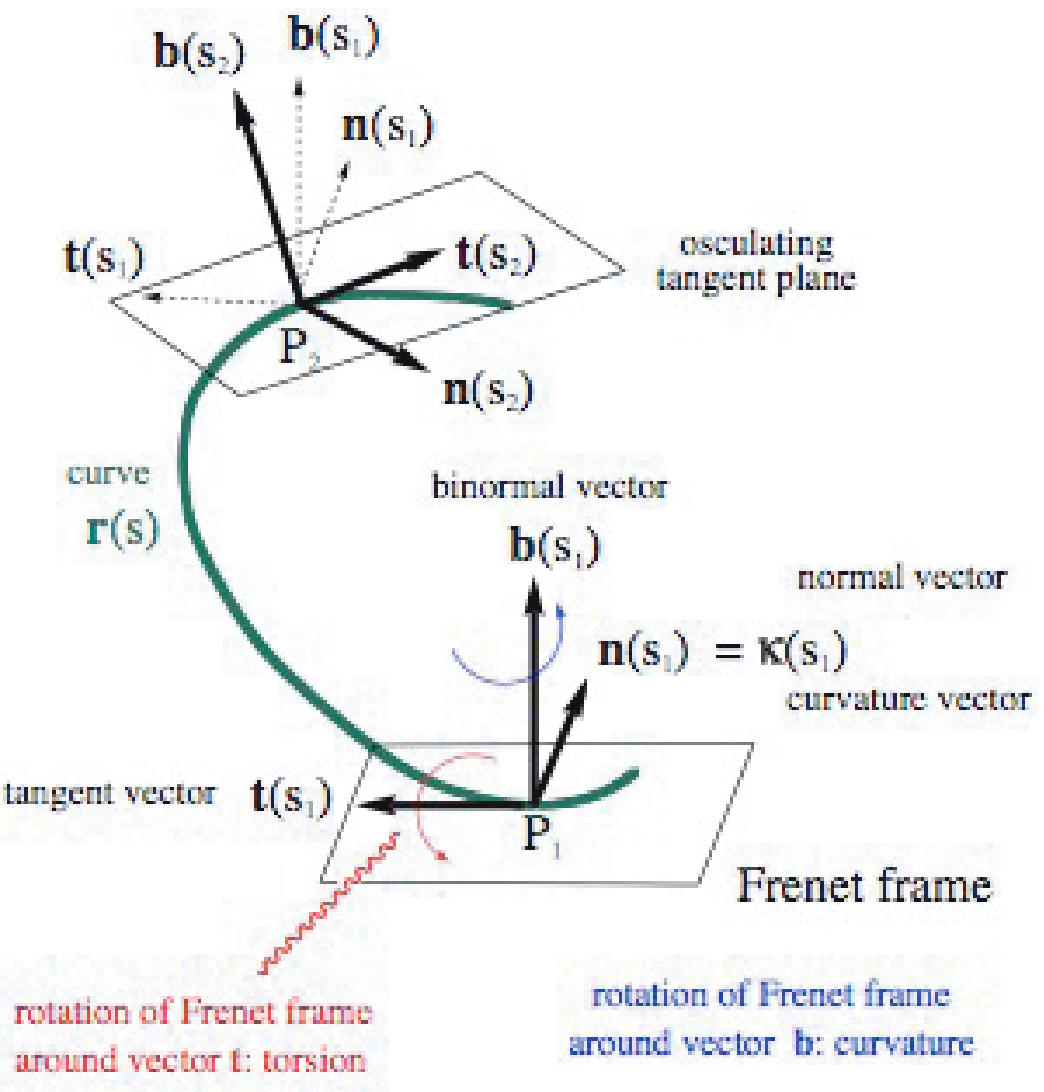}}

\noindent {\bf Figure 2}.   Schematic representation of tangent {\bf t}(s), normal {\bf n}(s), and binormal vector {\bf b}(s) vector of a Frenet frame at points
P$_1$ and P$_2$ of curve {\bf r}(s) presenting part of a left-handed helix. The normal vector is often also called curvature vector. Movement of the Frenet frame
leads to an osculating tangent plane,  a rotation at the tangent vector {\bf t}(s) (indicated by a red arrow), and a rotation at the binormal vector {\bf b}(s) (blue
arrow) thus specifying torsion and curvature, respectively. For point P$_2$, the old Frenet frame (dashed arrows) and the new Frenet one (bold arrows) are given, which
reveals that the torsion is negative according to a left-handed twist of  the Frenet frame around  {\bf t}.
\endinsert
\smallskip
For the purpose of converting the framework of bonds establishing the backbone into a smooth 3D line one has to answer two questions.
  a) What are the most suitable anchor points for the representation of the backbone? -  b) What type of
 spline function should be used? - The first question relates to the coarse graining strategy outlined in the introduction. If all
backbone atoms would be used as anchor points an excessively  detailed description of local conformational features with strongly oscillating scalar curvature and torsion
would result thus loosing the chance of describing non-local features of protein structure. Instead, we have to apply a first level of coarse graining by
presenting each residue along the backbone just by one anchor point. In this way, a smoother more global presentation of the backbone can be obtained
sacrificing unnecessary structural details such as the individual orientation of residue bonds in 3D-space.

There are various choices for the anchor points. These could involve the N or C(=O) atoms of the peptide linkages, alternatively the
C$_{\alpha}$ atoms, or any geometric point of a residue (center of mass, geometric center, etc.). After testing four
alternatives  (N, C(=O), C$_{\alpha}$, midpoints of C$_\alpha$ atom-pairs), we came to the conclusion that  protein structure is best presented  at the first level
of coarse graining by utilizing the C$_{\alpha}$ atoms as anchor points. The conformational flexibility of the
protein backbone is primarily determined by the C$_{\alpha}$ atoms and less by the atoms of the peptide linkage. Hence, any
description of protein conformation should reflect the position of the C$_{\alpha}$ atoms in 3D space.
This is in line with the fact that
from the early   $\phi - \psi$- Ramachandran plots [2] to more recent models of protein folding [25], geometrical descriptions of
the protein backbone  have repeatedly reverted to C$_{\alpha}$-based  representations. In the same spirit, graphical
representations of proteins indicating secondary structure preferentially utilize the C$_{\alpha}$ positions in 3D space. [26]

It has to be noted that the choice of the anchor points relates to some degree to the choice of the polynomial functions
selected to represent the protein backbone.  Of several possible spline functions that can be used for this purpose, the cubic spline
was found  most suitable, owing to its simplicity. It is uniquely defined by the requirement of smoothness at the anchor
points (C$_{\alpha}$ atoms) and by  two boundary conditions (see below). In addition, it corresponds to a curve with minimum
deformation energy. Other types of splines do not have this physical property; moreover, the higher-order splines
often require extra, nonphysical constraints, as they need more parameters to be fixed. Whereas the cubic spline functions have a
well-defined advantage when used with the C$_{\alpha}$ anchor points, their use in connection with the N, the carbonyl C
atoms or the mid-points of C$_\alpha$ atom-pairs does not lead to smooth backbone patterns in terms of curvature and torsion of
secondary  structures. 

The cubic spline functions used in this work employ {\it natural} boundary
conditions at either end, i.e., the curvature $\kappa$ is fixed to zero at the terminal C$_{\alpha}$ atoms of a protein. This
approximation is reasonable considering the high conformational flexibility of  the ends of most proteins, which are not
associated with a constant $\kappa$ value.  Test calculations (see Appendix I) revealed that the first and the last two  residues
of the protein backbone are affected by the boundary conditions and therefore they are excluded from the analysis.

The accuracy of the representation of the protein backbone
by a spline function  depends in turn on the accuracy of the coordinates supplied. Therefore, the sensitivity of the cubic
spline interpolation to any uncertainty in coordinates  has been quantified. The analysis reveals that at the first level of coarse graining
(with the C$_{\alpha}$ atoms as anchor points) calculated
$\kappa(s)$ and
$\tau(s)$ values  are more accurate than 0.1 {\AA}$^{-1}$ as long as the resolution {\it R} of the protein coordinates is $\le$ 2 {\AA}. 

The new {\it A}utomated {\it P}rotein {\it S}tructure
{\it A}nalysis (APSA) method is carried out  utilizing a  program that extracts the C$_\alpha$ coordinates  from a PDB file, [6] calculates the protein
backbone line, determines $\kappa(s)$ and $\tau(s)$, and then, utilizing a set of predefined rules coined from ideal secondary
structure patterns of $\kappa(s)$ and $\tau(s)$, automatically analyzes and identifies all secondary structural features of the
protein considered. Curvature values are always positive whereas torsion values can be both positive and negative. The sign of the scalar torsion is
defined  by the rotation of the binormal vector {\bf b} 
around the tangent vector of the curve. If it is clock-wise (along a right-handed screw), the torsion isÊ
positive; otherwise it is negative as in the case of the left-handed helix of Figure 2. Equation (2)  implicitly includes the direction of the rotation of the binormal
vector  and, hence, the sign ofÊ
the torsion. Given that APSA analyzes proteins from the first residue at the N-terminus, any change of sign of $\tau(s)$ correctly reflects any change of chirality of a
protein helix or any other structural unit.

In this work, APSA is used to investigate ideal secondary  structural units of proteins. Suitable references for the latter were constructed with the help of an
18-residue polyalanine helix and $\beta$-strand. For this purpose, ideal $\phi$ and $\psi$ angles of -57$^\circ$ and -47$^\circ$ for the $\alpha$-helix, -49$^\circ$ and
-26$^\circ$ for the $3_{10}$ helix, and  -57$^\circ$ and -70$^\circ$ for the $\pi$ helix were used.  A left handed $\alpha$-helix was constructed using the
ideal backbone angles  $\phi$ and
$\psi$ of 60$^\circ$  and 54$^\circ$.  An ideal $\beta$ strand is not planar as it is often sterically influenced by neighboring structures and therefore, it resembles a
twisted stretched  ribbon. This effect was modeled using
$\phi$ and $\psi$ angles of  -120$^\circ$ and 120$^\circ$. 

  The speed  and the reliability of APSA's
assignments were tested on a dataset of 20 proteins from the PDB [6] listed in Table 1.  Only X-ray structures having a
refinement of about 2 {\AA} or less were selected. Structure breaks and proteins with alternate locations provided for
C$_{\alpha}$ positions were  avoided though the $\kappa(s)$ and $\tau(s)$ patterns did not differ much in these 
cases. The set of proteins used for the APSA description contained different sized proteins with various lengths of helices and
$\beta$ sheets, connected by small and large sized loop regions.  The CATH classification system
[8] was followed making sure that the protein examples chosen spread over the main classes. 

\bigskip
\bigskip

\noindent {\bf 3. Analysis of Ideal Secondary Structures}
\smallskip

Table 1 contains the set of 20 proteins investigated in
this work and  summarizes the number of  secondary structures and residues in ideal conformational environments contained
in these proteins. In Table 2, Figure 3, and Figure 4,  suitable references of $\kappa(s)$ and  $\tau(s)$ values (and diagrams) of ideal secondary  structures formed by 
18-residue long polyalanines  (right-handed  ideal 3$_{10}$, $\alpha$-, and
$\pi$-helix in Figure 3a and Figure 4, left-handed  $\alpha$-helix in Figure 3b; $ ideal \beta$ strands in Figure 3c) are given.   
\bigskip 
\bigskip


\centerline{\epsfbox {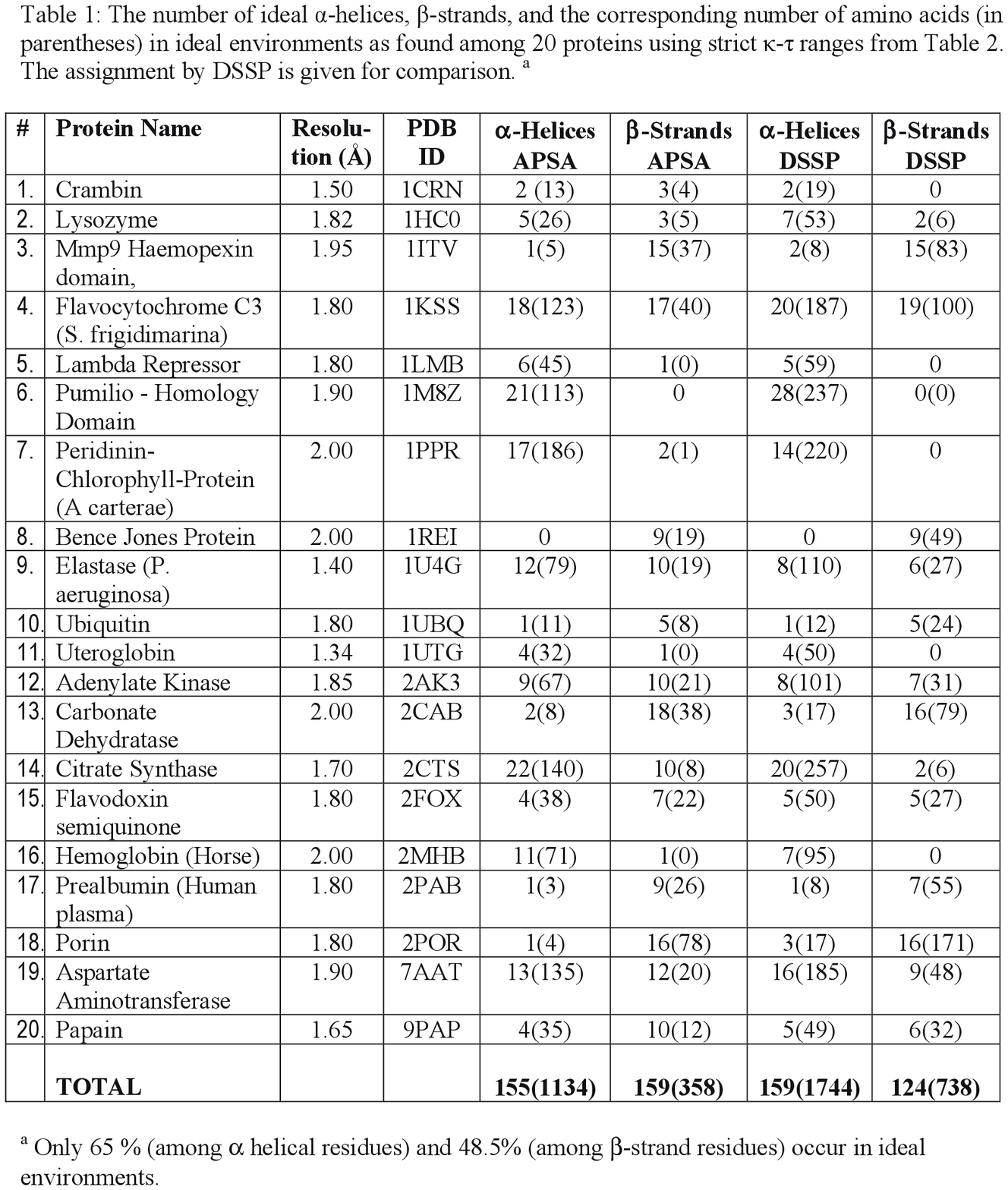}}


\epsfbox {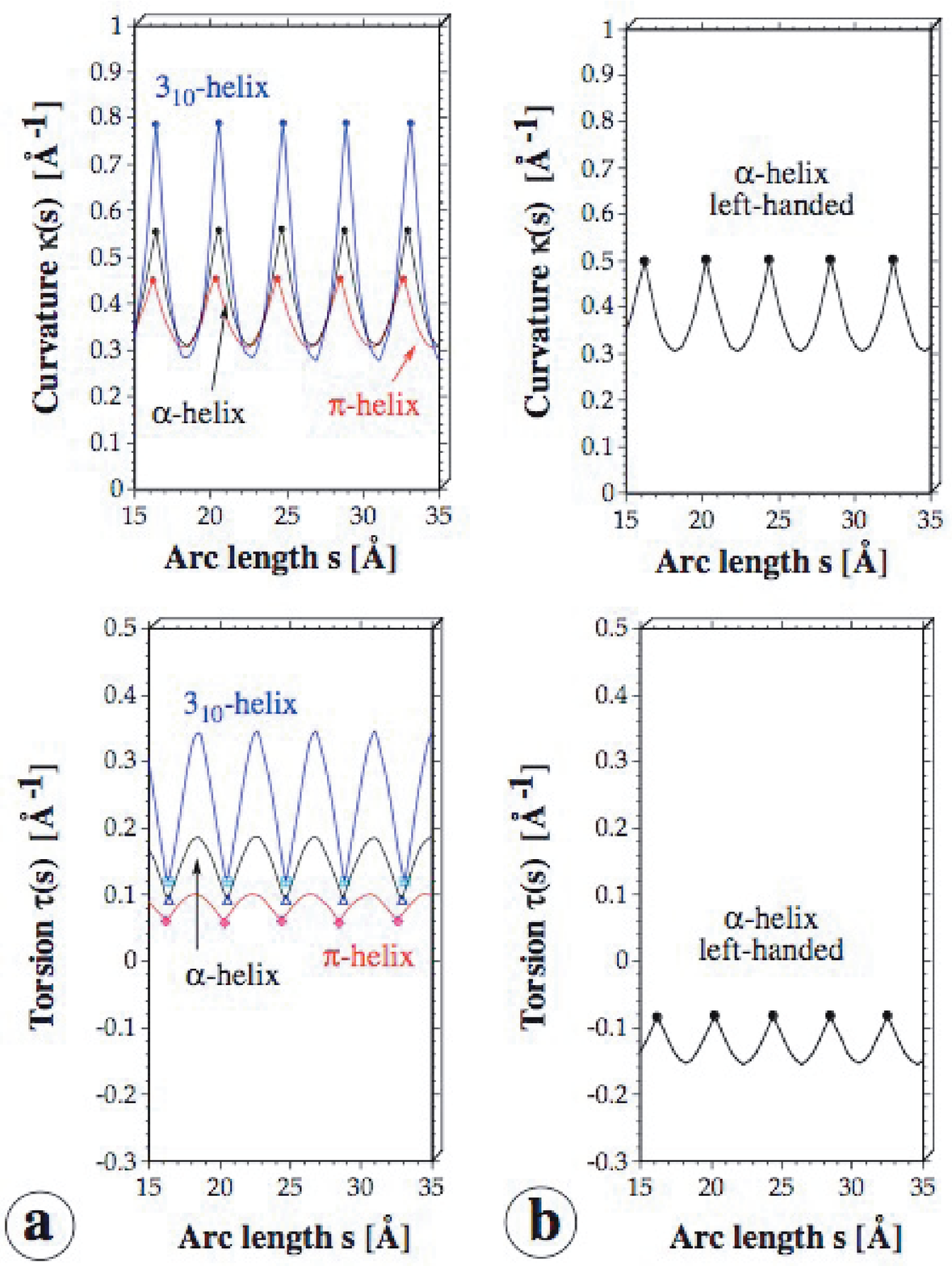}
\epsfbox {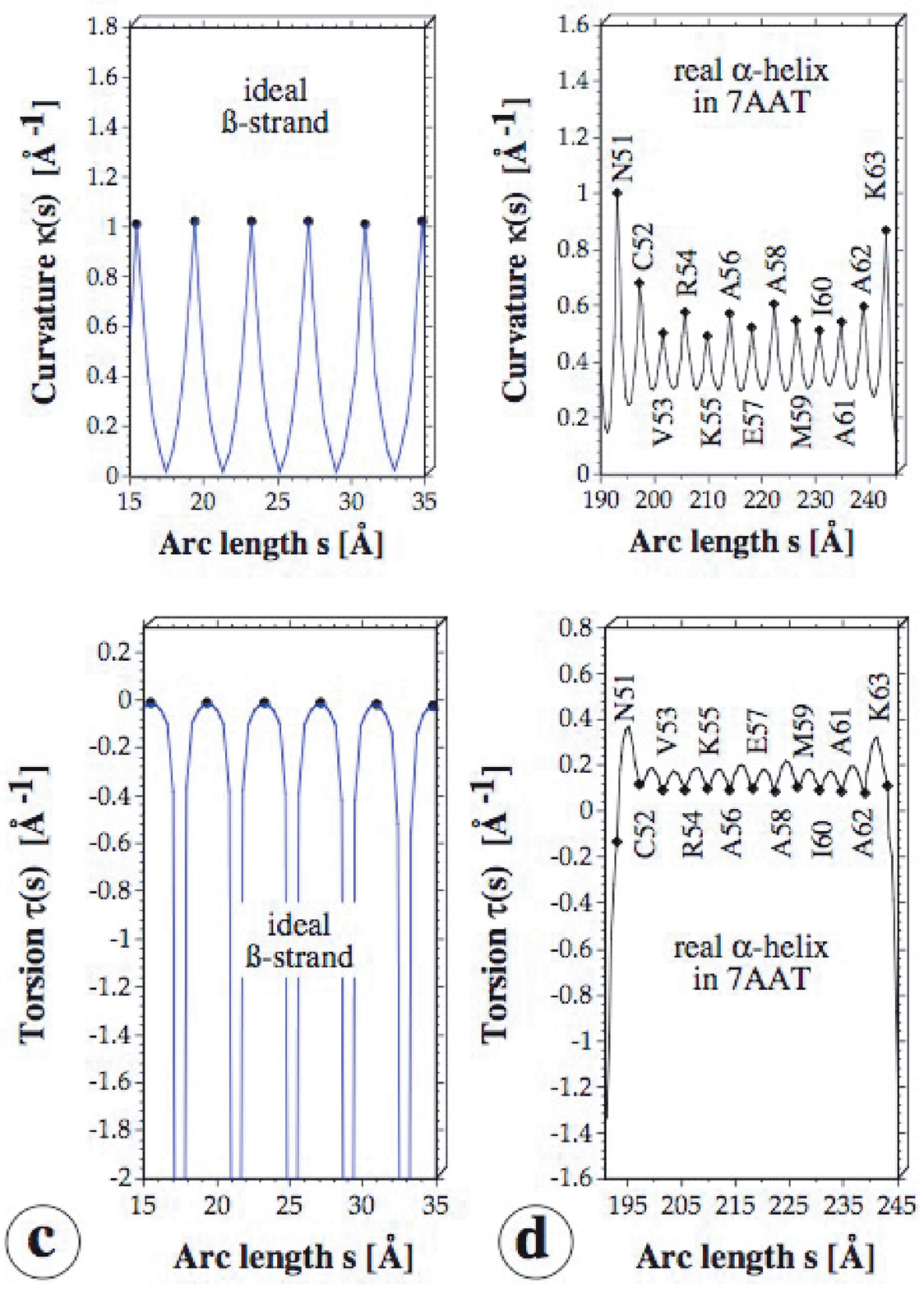}



\noindent {\bf Figure 3}.   Curvature diagrams $\kappa(s)$ (above) and torsion diagrams $\tau(s)$ (below) of (a) right-handed  ideal 3$_{10}$,
$\alpha$-, and
$\pi$-helix; (b) left-handed ideal $\alpha$-helix, (c) ideal $\beta$-strand; (d) natural $\alpha$-helix from N51 to K63 in protein 7AAT [6];  (g) natural $\beta$-strand
from T17 to K23 in protein 2SOD. [6]

\smallskip

An ideal helix  has constant curvature and torsion values; for example, for a
diameter of 1.0 {\AA},  $\kappa(s)$ = 0.5 {\AA}$^{-1}$. For the ideal 18-residue long polyalanine, the three types of
ideal helices have different $\kappa(s)$ and $\tau(s)$ patterns, which can be distinguished qualitatively and quantitatively (Figure 3a). The differences in
$\kappa(s)$ and
$\tau(s)$ values correspond to the differences in pitch and diameter of each helix. Since a 3$_{10}$ polyalanine helix has an i $\rightarrow$ i+3
H-bonding pattern (first and third residues are connected by a C=O $\cdots$ H-N bond), the diameter of a 3$_{10}$-helix is smaller than that of an $\alpha$-helix. 
Only 3 residues form a turn  (each residue corresponds to a 120$^\circ$ turn) giving a translation step of 2.0 {\AA} (along the helix axis for one loop).
Clearly, the tighter  3$_{10}$-helix must have larger curvature and torsion oscillations than the $\alpha$-helix (Figure 3a). For the  $\alpha$-helix, the
 i
$\rightarrow$ i+4 H-bonding pattern, results in 3.6 residues  per turn where a residue leads to a 100$^\circ$ turn and the translation step is 1.5 {\AA}. 
The   $\pi$-helix is characterized by the i
$\rightarrow$ i+5 H-bonding pattern, 4.1 residues per turn, a 87$^\circ$ turn per residue, and a translation step of 1.15 {\AA}. Hence, oscillations in
curvature and torsion are smallest for the $\pi$-helix and intermediate for the $\alpha$-helix.
 Considering the reduction in the diameters and the transversion step
of 3$_{10}$, $\alpha$-, and $\pi$-helix in that order, the amplitude of oscillation in the curvature and torsion must decrease as seen from
the $\kappa(s)$ and $\tau(s)$ diagrams of Figure 3a.

The functions $\kappa(s)$ and $\tau(s)$ oscillate about an average value, which is an
innate property of the helix in question: 0.46 and 0.24 {\AA}$^{-1}$ in case of the 3$_{10}$-helix indicating strong curving and torsion of the backbone, 0.40 and
0.15 {\AA}$^{-1}$ in case of the $\alpha$ helix (intermediate curvature and torsion), 0.36 and 0.08 {\AA}$^{-1}$ in case of the
$\pi$-helix (weak curvature and torsion). The curvature maxima coincide with the position of the  C$_{\alpha}$ atoms (dots in
Figure 3a,
$\kappa(s)$ = 0.81, 0.55, and 0.45 {\AA}$^{-1}$, respectively; Table 2) because they are the points of strong bending. For an
$\alpha$ helix, these points lie on the vertices of the approximate 'square' formed when the helix is seen end-on. The plane of the amide bond enforces
linearity to the region in between the C$_{\alpha}$ points resulting in curvature minima (Figure 3a, $\kappa(s)$ = 0.28, 0.30. and 0.30 {\AA}$^{-1}$, Table
2). 



\centerline{\epsfbox {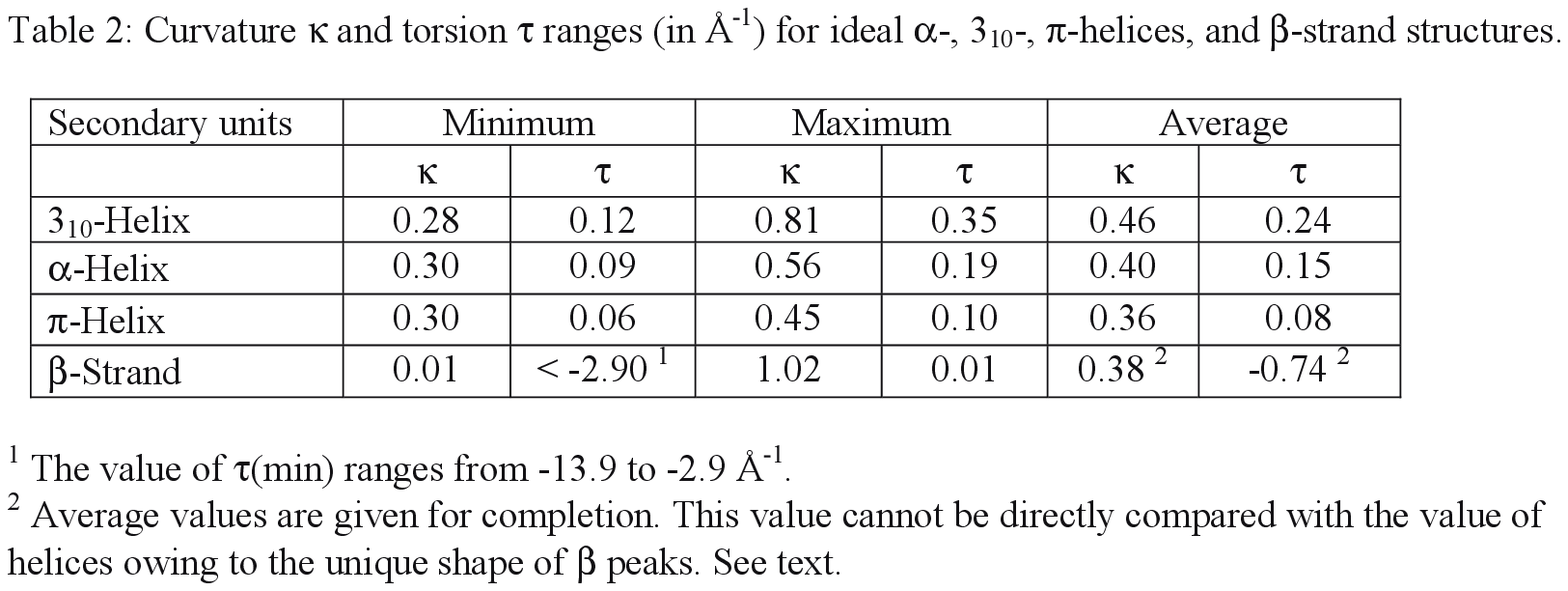}}
\bigskip
\bigskip

\centerline{\epsfbox {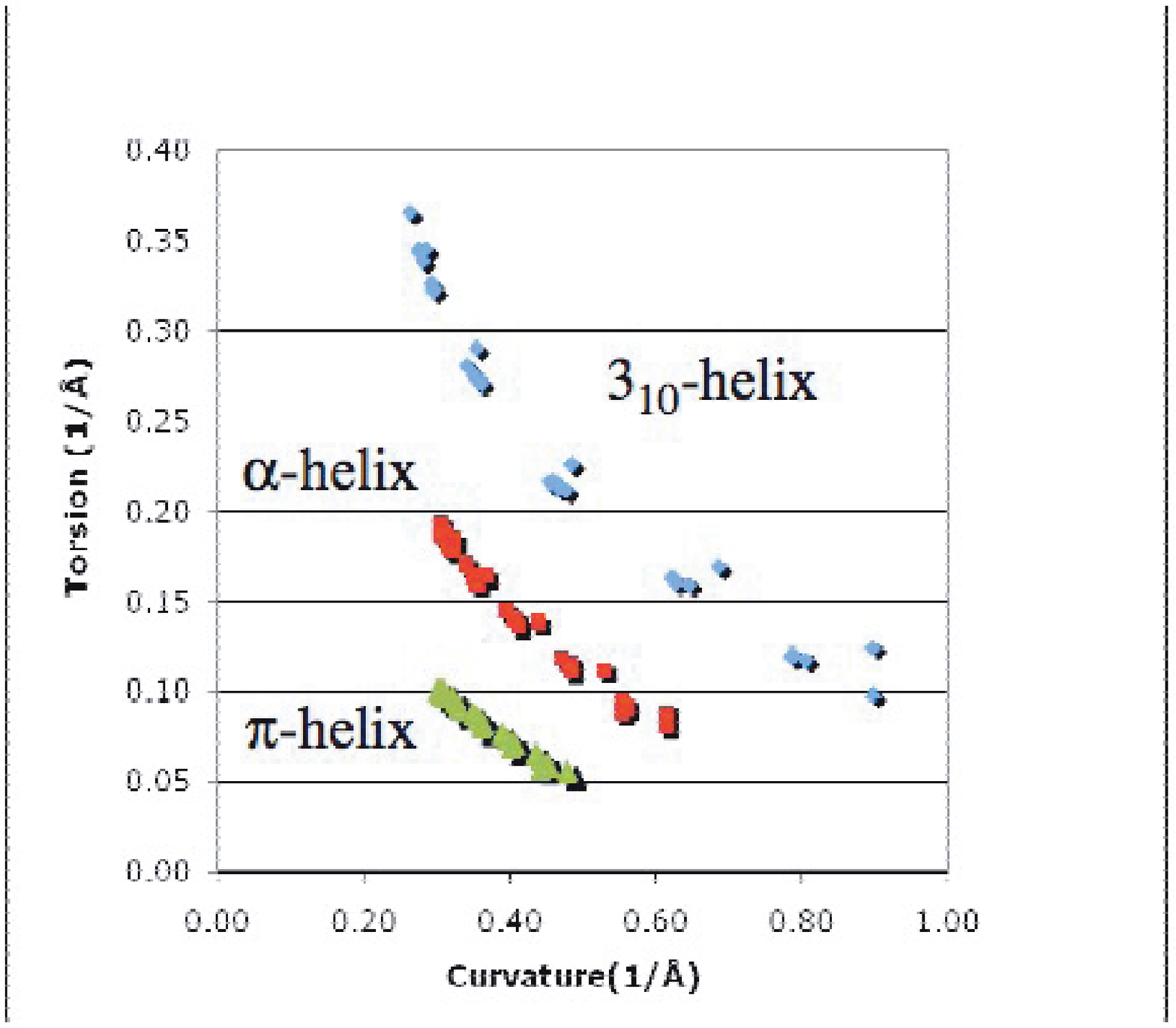}}
\smallskip

\noindent {\bf Figure 4}.  The torsion $\tau$ is given at 20 points between two consecutive C$_\alpha$ anchor points of the helix backbone in dependence of the
corresponding curvature values $\kappa$ for an ideal 3$_{10}$helix (blue points), an ideal $\alpha$-helix (red points), and an ideal $\pi$-helix (green points).  

\smallskip


A similar
reduction is found for the torsion maxima of the 3 helices (0.35, 0.19, 0.10 {\AA}$^{-1}$, Table 2) reflecting the decrease in the transversion steps from 2 to 1.5, and
1.15 {\AA}, respectively (Figure 3a,
$\tau(s)$). The strongest torsion is found in the peptide units because these point in the direction of the helix axis giving 
maximum contribution to the rise per amino acid. At the C$_{\alpha}$ atoms, the torsion is at its minimum (0.12, 0.09, and 0.06  {\AA}$^{-1}$, Table 2, see also $\tau(s)$
in Figure 3a).

In addition to the $\kappa(s)$ and $\tau(s)$ diagrams, 
the characteristic differences between the three helices can  be observed in a $\kappa$-$\tau$ diagram  (Figure 4). In this case,
 $\kappa(s)$ and $\tau(s)$ are calculated for 20 points along the spline for every  two successive C$_{\alpha}$ atoms and  are presented
as function $\kappa = f(\tau)$. For  3$_{10}$, $\alpha$, and $\pi$ helix, three different seggregations of points are obtained  where the 3$_{10}$ helix has again
the largest and the $\pi$ helix the smallest $\kappa$-$\tau$ values. The lack of overlap between the group of points for each helix confirms that  unique values of
$\kappa(s)$ are obtained for the corresponding values of $\tau(s)$ throughout the backbone of each helix. The sharp turning of a 3$_{10}$ helix leads to stronger
oscillations in the curvature and torsion (Figure 3a), and corresponds to the wider spread of  points ( Figure 4). The $\pi$-helices, having a more
relaxed structure caused by more amino acids along each  turn, show the least dispersion of points.  The $\alpha$ helix takes the intermediate
position as already discussed in connection with Figure 3a.

 For a right-handed helix, torsion values are positive whereas they are negative for a
left-handed helix: average $\tau(s)$ = 0.15  and -0.15 {\AA}$^{-1}$, respectively (see Figures 3a and 3b). In this way, the chirality of the helix is
directly reflected by the torsion values of the protein backbone determined by APSA.

Ideal $\beta$-strands are regular pleats of the protein backbone, rotated along the strand axis  where
the majority of $\beta$-strands is twisted in a left-handed fashion. 
This is reflected in the typical patterns of the $\kappa(s)$ and $\tau(s)$ diagrams for the ideal $\beta$-strand (Figures 3c). 
$\beta$-Strands can be considered as the sterically stable conformation of a  helix that has two amino acids per turn.  
The C$_{\alpha}$ atoms lead to strong curving of the backbone line (high  $\kappa(s)$ values) but do not contribute
much to the overall rise of the strand ($\tau(s) \approx$ 0) whereas the region in-between the C$_{\alpha}$ atoms is relatively straight ($\kappa(s) \approx$ 0), but
responsible for stretching of the strand. The turning of the spline is left-handed resulting in the strong negative torsion minima.
  Though the average $\kappa(s)$ value of 0.38  {\AA}$^{-1}$) (Table 2) of the ideal $\beta$-strand is lower than that of the $\alpha$ helix (0.40 {\AA}$^{-1}$), it should
be noted that the shape and height of the $\beta$ peaks are  different. Its height, given by the difference between the maximum and minimum
of each peak, is 1.02 - 0.01 = 1.01 {\AA}$^{-1}$ whereas for the $\alpha$ helix it is 0.55 - 0.30 = 0.25 {\AA}$^{-1}$.

 Figures 3d and 3e show an $\alpha$ helix and a $\beta$ strand taken from a segment of natural proteins. The former structure belongs to region N51 to K63 in aspartate
aminotransferase (7AAT) and reveals  variations in the curvature and torsion maxima that indicate deformations of the helix. The $\beta$-strand from region T17 to K23 in
superoxide dismutase (2SOD) shows even more fluctuations in curvature and torsion than the ideal structure in Figure 3c. In both cases, the ends of the structures are
not as perfect as their ideal counterparts. This is because  they continue into  neighboring turns which are non-planar. These examples emphasize that
the curvature and torsion patterns are sensitive to any deformation of a real secondary structural unit. In addition, they make it possible to exactly describe entries
and exits to these units, which  opens up new prospects for an automated secondary structure recognition by APSA.

The three standard helices (3$_{10}$,
$\alpha$-, $\pi$-helix) and the $\beta$-strands can be distinguished by their average and extreme curvature and
torsion values, as summarized in Table 2. The APSA method has been programmed to do this by utilizing the characteristic values for each
structure (Table 2), whether it is in an ideally constructed polyalanine or a real protein, as will be discussed in the
following sub-section. Since variation in these values is observed for $\beta$ strands, a value of $<$ -2.9 {\AA}$^{-1}$ is used to define the
$\tau(s)$ minimum. The various residues in the protein backbone can be appropriately determined as part of a helix or
$\beta$-strand by defining lower and an upper limits of $\kappa(s)$ and $\tau(s)$ values in form of a {\it $\kappa-\tau$ window} for each specific
secondary structure. 
\medskip

\noindent {\bf 3.1 Distribution of amino acids in ideal conformational environments}
\smallskip

 Literature is replete with comparisons of the results of various secondary structure assignment
methods. It has been shown [1] that automated methods such as DSSP [4], DEFINE [12], and P-curve [13] agree in only  63 \% of residues among
helices, $\beta$-strands and non-regular structures. Additionally, it has been shown that $\beta$-strands showed greater disparity owing
to differing definitions. [1] A recent automated assignment based on structure comparison involving STRIDE, [11] DSSP, [4] and the PDB [6] using
fuzzy logic [27] revealed that the middle regions of helices show maximum agreement whereas the ends are the disputed regions. In the wake of such conflicts, it is useful
to evaluate the extent of helical and $\beta$-strand regions that are strictly ideal in conformation. Also, rigid helices in proteins are significant in protein folding,
protein-ligand and protein-protein interactions where they function as scaffolds onto which the more variable regions such as caps interact.
APSA can be used to determine how many amino acids in a protein contribute to ideal conformations in secondary structures. This
number reflects the conformational stability of those regions, as opposed to conformational flexibility where folding and long
range interactions lead to distortions. 

APSA, as discussed under section 2, is non-local and continuous in its presentation of
the protein backbone and therefore it recognizes conformations reflecting global trends. An amino acid having an ideal conformation will be
recognized as {\it ideal} only if it occurs in an ideal {\it environment} with a distinctive pattern in $\kappa$ and $\tau$ that reflect an
ideal collaborative conformation. One localized $\alpha$-helical amino acid
(having the ideal $\phi$ and $\psi$ value) in a region that is significantly distorted cannot be counted as an ideal $\alpha$-helical residue. It takes an
accumulation of 3 to 4 such ideal amino acids to create a secondary structure that has the ideal  $\alpha$-helical pitch and diameter. It is this
type of pattern that APSA reads from the $\kappa(s)$ and $\tau(s)$ diagrams. The influence of
neighboring amino acids on the $\kappa$ and $\tau$ values of the amino acid in question, however, depends on the overall shape of the secondary structure  being formed by the
simplified backbone. Though it is a universal rule that this type of influence decreases with distance, drastic changes in
conformation up to two C$_\alpha$ positions away can affect the $\kappa$ and $\tau$ values of the amino acid in question. 

Table 1 lists for 20 proteins  the number of amino acids occurring in ideal environments. Results are
compared with DSSP [4] as it is a widely used hydrogen bond-based method capable of recognizing $\alpha$, 3$_{10}$ and $\pi$ helices, bends, turns,
$\beta$-strands and isolated $\beta$-bridges among amino acids that participate in hydrogen bonding. In the current comparison only $\alpha$-helices and
$\beta$-strands longer than 2 amino acids were considered by APSA.

\topinsert


\centerline{\epsfbox {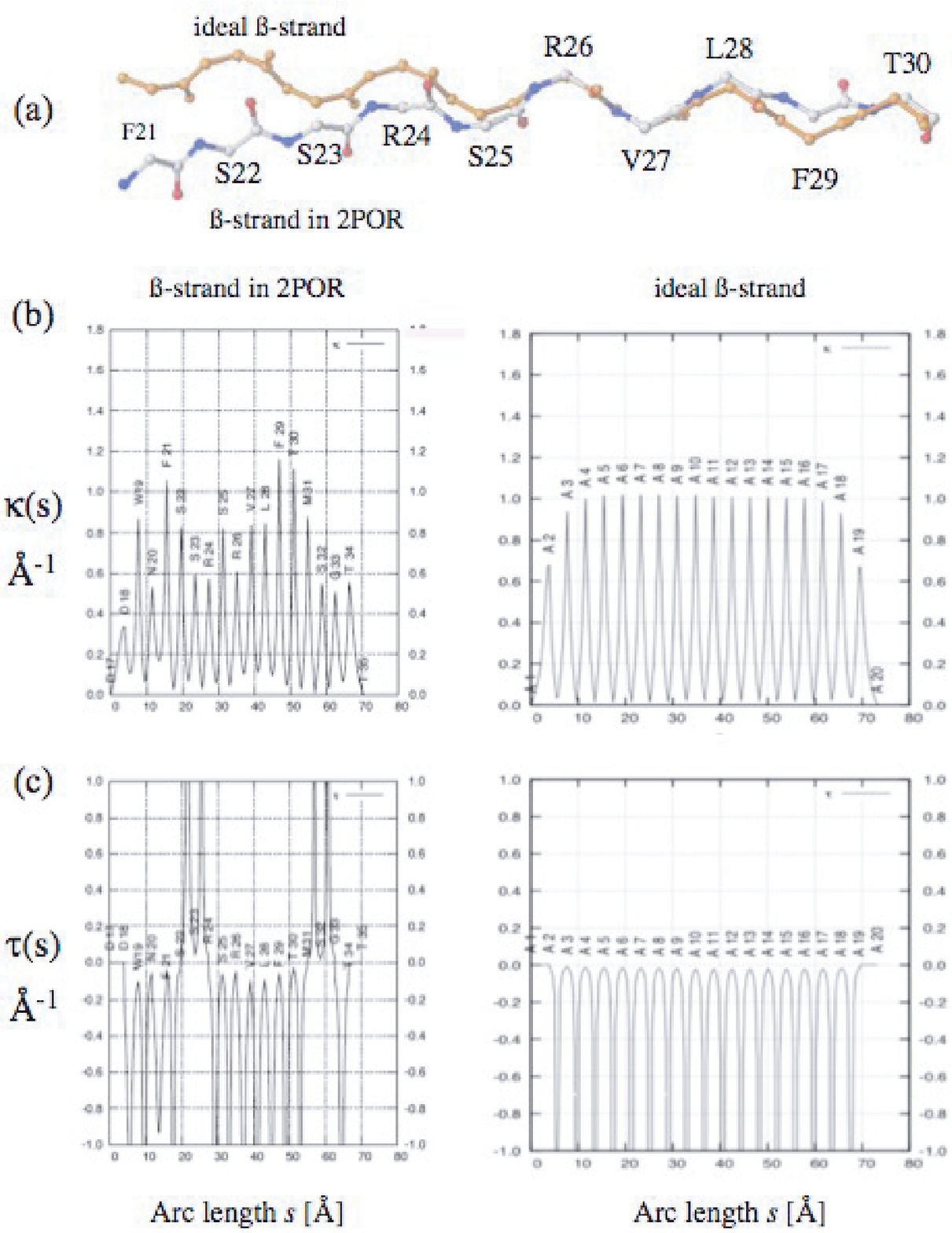}}

\noindent {\bf Figure 5}.   (a) A Ball\&Stick representation of the backbone region W19 to G33 in 2POR (a porin; C: white; N: blue; O: red color), which deviates from a
strictly ideal $\beta$-strand (brown color). The position of the carbonyl oxygen (red atoms) indicates torsion of the backbone besides significant backbone
curvature. (b) Curvature and (c) torsion diagrams, $\kappa(s)$ and $\tau(s)$, quantify deviations from the ideal $\beta$-strand.
\endinsert
\smallskip  

Among the 20 proteins, 155 helices are recognized comparable to the 159 identified by DSSP (Table 1).  However,
the number of residues possessing an ideal $\alpha$-helical conformation  is lower than the DSSP
number of helical residues . Out of DSSP's 1744 helical amino acids, only  65 $\%$ occurred in ideal
helical environments. In some proteins such as  2FOX, 2POR, 7AAT,  etc., a few helices not being recognized are not $\alpha$-helical in shape
whereas in other proteins such as 1M8Z short helices with one or two ideal residues escape recording as the rest of the helix is
distorted beyond recognition. In protein 1HC0, there is great disparity between the difference in the number of structures (5 vs 7 helices)
and the ideal amino acids (26 vs 53) indicating that the protein contains helical structures of intermediate conformation between complete
distortion and the ideal $\alpha$-helix. Though APSA recognizes in some cases (1PR, 1U4G, etc.)  more helices than DSSP, this does not proportionally increase the number
of ideal residues. For example for 2CTS, just 140 (54 \%)  out of DSSP's 257 helices are ideal. A detailed account of the various distortions found in helices with
 quantification of these distorted structures will be the topic of following works. [28]

The $\beta$-strands recognized by APSA significantly outnumbers those recognized by DSSP. The disparity is because, being a geometry-based method, APSA
can analyze the conformation of all amino acids in the protein and does not require the presence of the $\beta$-ladder.  Consequently, APSA identifies more
$\beta$-strands, but at the same time all identified $\beta$-strands are considerably shorter thus reflecting APSA's sensitivity with regard to any distortion.
In total, the percentage of  residues in an ideal $\beta$-environment is just 48.5. The inference that can be
obtained directly from this analysis is that conformationally, the strands participating in a $\beta$-sheet are not as rigid as helices. The same concept has been
discussed in early literature, indirectly, in terms of the inconsistency encountered when calculating and modeling hydrogen bonds
into $\beta$-ladders [20].

The distortion of $\beta$-strands from ideal structure can be made visible by the superposition of one distorted
example on an ideal structure as shown in Figure 5a. 2POR is a porin with the $\beta$-barrel architecture having 16 $\beta$-strands with
roughly 46 $\%$ of the residues being ideal. The backbone of a DSSP assigned $\beta$-strand from W19 to G33 
is an example of a deviation from the ideal structure as a result of an overall curving of the strand and a simultaneous
 rotation of the backbone (as seen from the staggering of the red carbonyl oxygen atoms). The $\kappa(s)$ and $\tau(s)$ diagrams (Figures 5b and 5c) reveal these
differences clearly (e.g., changes from left to right, back to left, then to right, and finally a left-handed torsion of the strand accompanied by varying curving) and
provide a basis for  discussing distortions without the need of a 3D-comparison (as done in Figure 5a). [28a] 
\medskip

\noindent {\bf 3.2  Analysis of non-regular structures}
\smallskip

	From the above discussions, it is clear that the geometry-based APSA method can efficiently analyze regular
repeating secondary structures. But turns are more variable and, despite several approaches used to describe them, [15]
remain so far elusive to any form of classification. The term loop refers to the regions of the proteins not recognized as turns by the
respective classification system used for structure description, and these regions comprise about 50 \% of the protein. [15] APSA's
spline fitting procedure connects all the C$_\alpha$ points in the protein and is solely based on conformation excluding the need for
any other information but the C$_\alpha$ positions. Turns have been considered in literature [15] as a combination of helical and
extended geometries. Given that protein structures are described in both ideal and real forms by APSA (Table 1), the description of turns should
thus reflect the exact 3D-shape of a turn and be included in the $\kappa(s)$ and $\tau(s)$ diagrams of a complete protein, which will be tested in the following. 
\bigskip

\topinsert
\centerline{\epsfbox {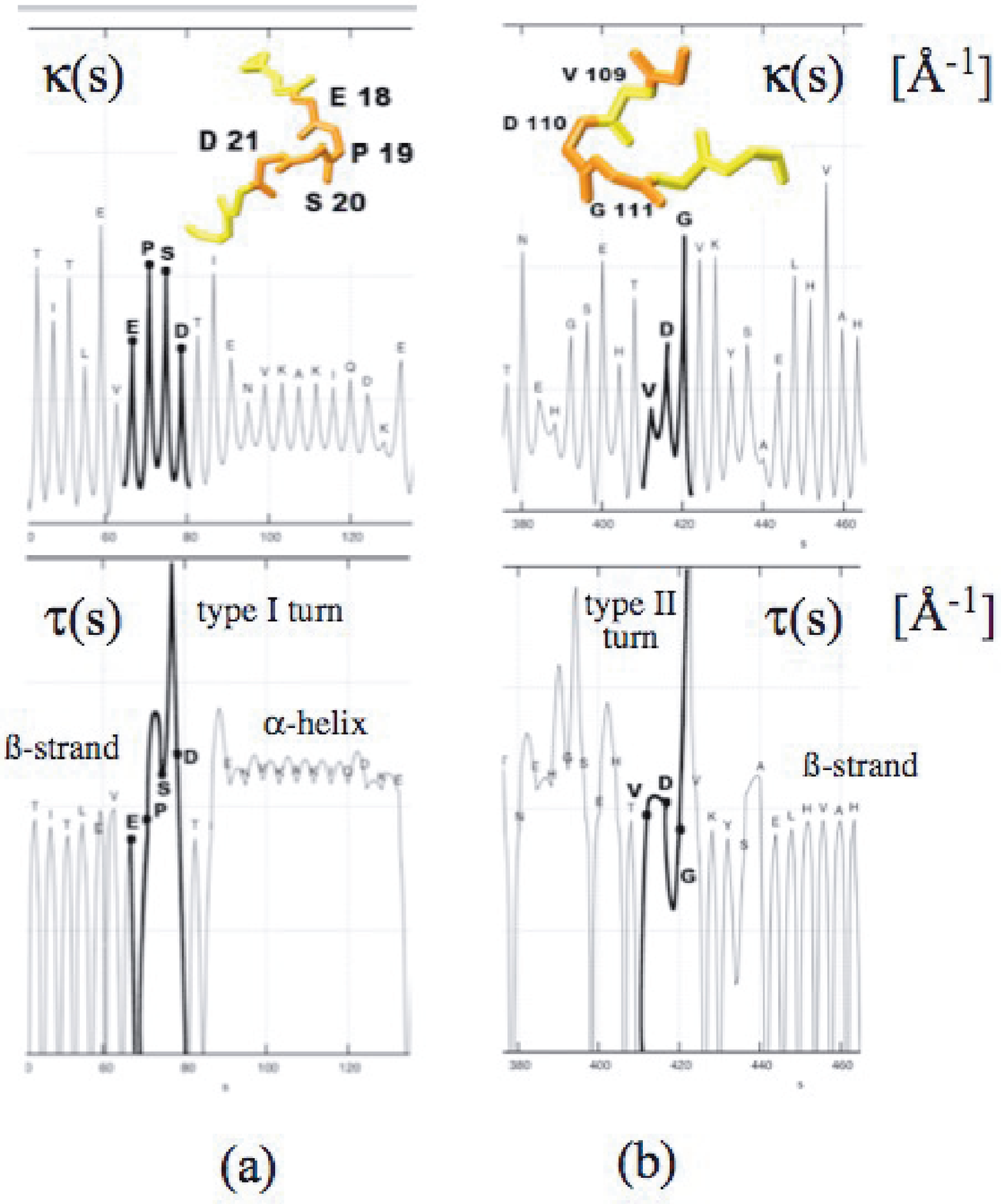}}

\noindent {\bf Figure 6}.  Calculated $\kappa(s)$ and $\tau(s)$ diagrams of (a) a type I turn  as
occurring in ubiquitin (1UBQ) region E18-D21, and (b) a type II turn from carbonic anhydrase form B  (2CAB) region V109-V112.  Insets: the backbone is colored
orange in the turn regions.
\endinsert
\bigskip

 Figure 6a shows   a type I turn in the backbone of ubiquitin (1UBQ, amino acids 18 to 21; colored orange in the
inset). The $\kappa(s)$ and $\tau(s)$ diagrams of this region have the peaks of the amino acids
of the turn highlighted in bold. The pattern of the turn is not  a regular repeating secondary
structure, however, the helical $\tau$-peak at proline 19 and the $\beta$-like $\tau$-peaks at E18, S20 and D 21 are recognizable. It is
evident from the backbone representation  that the proline is mainly responsible for the turning of the
backbone  and that the backbone going 'downwards' in the inset picture of Figure 6 raises  a little at serine 20 before it
starts going down again corresponding to the sign changes in torsion. An inspection of the other regions of the protein displayed in
$\kappa(s)$ and $\tau(s)$ diagrams shows that the turn occurs between a regular $\beta$-strand to the left and a regular $\alpha$-helix
to the right. Similarly, in Figure 6b, the type II turn from amino acid V109 to G111 (peaks in bold in the $\kappa$ and $\tau$ diagrams) occurs
mainly at the aspartate residue at D110 where the torsion peak is helical. In this case however, in contrast to the turn example
in Figure 6a, the helix is left handed as shown by the negative torsion peak at G111 and by the direction of the turn in the backbone
picture. Two turn regions (both colored orange) can be recognized in the inset of Figure 6b, and the part of the $\tau(s)$ diagram preceding the V109-G111 turn shows
neither helix nor strand pattern. From the 
$\kappa(s)$ and $\tau(s)$ diagrams, it can be seen that the type II turn is followed by a $\beta$-strand distorted in the beginning. 

Due the fact that the  $\kappa(s)$ and $\tau(s)$ diagrams of the type shown in Figure 6 are  detailed containing all the conformational information needed, 
similarity between proteins or parts of proteins  in 3D can easily be assessed. To address this issue, 6 segments from the proteins 1JZB (variant 2 scorpion
toxin), 2PAB chains A \& B (prealbumin), 2TPI (trypsinogen), 5PTI (bovin pancreatic trypsin inhibitor), and 7RXN (rubredoxin), known to
contain the same turn, were analyzed. The $\kappa(s)$ and $\tau(s)$ diagrams of these segments cut out of the respective proteins are
shown in Figure 7 with a representative backbone rendering from 5PTI (inset of Figure 7).  It can be seen that the $\kappa(s)$ and $\tau(s)$ diagrams of
the turn regions are similar in all proteins both qualitatively and, to a large extent, even quantitatively. 

\smallskip


\centerline{\epsfbox {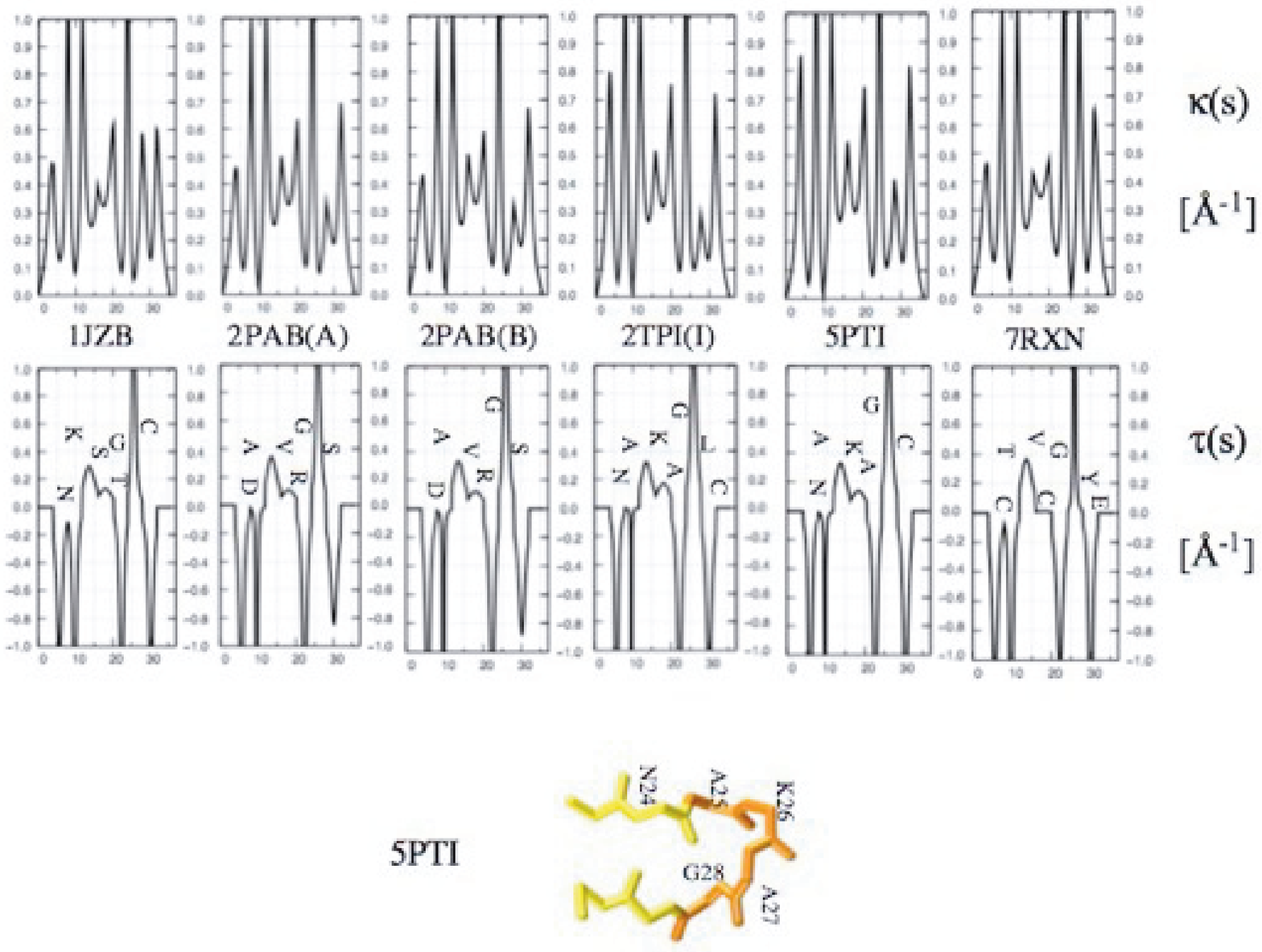}}

\noindent {\bf Figure 7}.  Similar conformational regions have similar $\kappa(s)$ and $\tau(s)$ patterns. The $\kappa$(s)- (above) and $\tau$(s)-patterns (below) of
a turn region (7 residues long)  that occurs in six different  proteins (PDB notation [6]: 1JZB, 2PAB(A), 2PAB(B), 2TPI(I), 5PTI(I), 7RXN) is shown. The inset gives
the turn region for the backbone rendering of protein 5pti.  
\smallskip

Of the 7 residues shown in each turn (Figure 7), the first is always $\beta$-like (compare with Figure 3c) and extends up to the C$_{\alpha}$
point of the second ($\kappa(s)$ diagram), after which it turns helical as seen from the high minima (and short peak height) of
$\kappa(s)$  and the helix-like shape of the $\tau(s)$ peak. This shape continues through the following residue up to the
C$_\alpha$ of the fourth. The fifth and the sixth residues are again extended, but pointing up and down as shown by the backbone
rendering and by the $\tau(s)$ sign changes. The C$_{\alpha}$ atom  of the fifth residue (a glycine in all
cases shown) is at the point of a sharp change in backbone orientation as shown by the strong $\kappa(s)$ $>$ 1.4 and by the $\tau(s)$
changing sign from strongly negative to high positive values. The turn is on the whole right-handed as revealed by the positive torsion
values in the helical region (residues 2 and 3; Figure 7).

\bigskip
\bigskip

\noindent {\bf 4. Comparison of APSA with Related Protein Structure Description Methods}
\smallskip

The original idea of developing APSA was based on the work with the Unified Reaction Valley Approach (URVA)  of E. Kraka and D.
Cremer. [29,30] There, the  reaction path of a chemical reaction is presented by a curve in multi-dimensional space, which is
characterized by arc length, curvature, and torsion where the latter quantities are related to energy transfer and energy
dissipation during the chemical reaction. The work with URVA led to the question whether protein folding can be also described 
by a conformational path, in which the protein backbone changes its shape in a typical way. From there, it was just a small step
to represent the backbone itself as a smooth  line in 3D space, which can be characterized by arc length,
curvature, and torsion.

Study of the literature revealed that related ideas have been pursued by   other authors although  mathematical procedures and 
 objectives of previous work differ significantly from those of APSA. In this section, we will shortly  compare previous studies
based on similar ideas of describing protein structure with the work presented here.

In 1978,  Rackovsky and Scheraga~[31] suggested a discrete analogue to APSA to evaluate $\kappa$ and $\tau$ at the C$_{\alpha}$
positions.  The protein backbone was represented by a set of "equidistant" points in space (C$_{\alpha}$ atoms) connected by
virtual bonds. The four points, $i-1,\dots,i+2$, form an elementary unit at point $i$ which defines $\kappa_i$ and
$\tau_i$. The graphical representations of $\kappa$ vs $\tau$ were  used for identification of structural elements and diagrams giving $\kappa$ or
$\tau$ in dependence of the residue number  were employed for protein comparison. This method conforms with the local
properties of curvature and torsion. The  bond angles and dihedral angles defined by the virtual bonds are related to the traditional
angles $\phi$ and $\psi$.  The virtual bonds, of course, differ significantly from the protein backbone  so that 
$\kappa(s)$ or $\tau(s)$ patterns identifying structural units of a protein can no longer be obtained. Hence, the method of 
Rackovsky and Scheraga~[31] is a forerunner of APSA based on a  representation of the protein backbone by a set of discrete points, which leads 
to shortcomings when analyzing protein structure.  

Louie and Somorjai~[32,33] considered the native conformation of a protein as the
collection of minimal surfaces, helicoids and catenoids, linked by turns and irregular coils, which contain mainly
polar groups and are exposed to solvent. The backbone is defined as a geodesic curve on the minimum energy surface. Only
helices and $\beta$-strands can be  analyzed in this way, whereas an analysis of "non-regular" residues (belonging to turns,
 loops, etc.) cannot be performed. In this approach, curvature
$\kappa$ and torsion $\tau$ are uniformly approximated by  step-wise functions corresponding to discrete values at C$_\alpha$
atoms. The fitting produces a sequential overlap of all portions of the
protein. Recognition of helices is based on  the comparison with an average cylinder of given radius and pitch. The
bends are recognized as changes in axial angles of the fitting helices, whereas turns are recognized when three or more
consecutive changes of axial angles are each greater than $40^\circ$.  The method cannot generally be applied and is best
suited for helix recognition.  

Soumpasis and Strahm [34] also considered describing the protein backbone by curvature and torsion obtained from polygonal
non-differential paths defined by C$_\alpha$ positions. At each vertex, curvature $\kappa$ is defined from the circumscribed circle of a
triangle spanned by three sequential C$_\alpha$ atoms. Torsion  
$\tau$ is defined in a similar way using a tetrahedron spanned by four C$_\alpha$ atoms. The formulas for calculating 
$\kappa$ and $\tau$ are
expressed in terms of Caley-Menger determinants. In this approach, the definition of $\tau$  suffered from an ambiguity and,
therefore it had to be augmented by an extra condition, which leads to a complex torsion with real and imaginary part. Using
this approach the authors obtained specific profiles characterizing secondary and supersecondary structure. [34] One of the
drawbacks of the Soumpasis-Strahm approach is that it does not produce a clear and easy-to-analyze picture of the backbone structure. 

Hausrath and Goriely suggested a continuous representation of  proteins from curvature profiles, [35] which were determined in a way similar to that of 
Louie and Somorjai.~[32,33] Small segments of the protein were described by a piece of a helix, which was then used for the calculation of curvature and torsion.
The latter obtained a stepwise character without specific patterns. Curvature and torsion were used for the calculation of amino acid atom coordinates 
in the local coordinate system. The averaged values of these coordinates were used to reconstruct the  protein conformation for similarity comparisons.

Sklenar, Etchebest, and Laverey~[36] suggested a method for smoothing the protein backbone. Each residue is assigned a local
helical axis system defined by a least-square-fit. The latter takes into account the
differences in the axis systems of the nearest neighbors. Local variations of atomic coordinates are "smoothed" out
and a more global picture can be obtained. Again, only a discrete set of points can be produced to represent the protein backbone. Differing orientations of helices
are described by the method. The regions with regular secondary structures (helices and $\beta$-sheets) are seen as
straight segments of a line, which are linked by curved segments corresponding to the non-regular conformations. This however leads to the loss of
 a visible difference between helices and $\beta$-sheets. Apart from this, structural details along the protein
backbone are lost during the smoothing process. The method contains several parameters, which  cannot be directly related to $\phi$ and
$\psi$ values or to protein conformation in general.  

Zhi and co-workers~[37]  suggested a smoothing technique for the protein backbone by
averaging C$_\alpha$ positions in a seven-residue window. Chain fragments that remain straight after  smoothing
are denoted as generalized secondary structure elements. The main characteristic is the turning angle along the
smoothed  backbone. Analysis and comparison of protein structures can be carried out by aligning the arrays of the angles. Though
this approach gives a more global view, it is unable to differentiate between secondary structural units.  Can and Wang~[38]
estimated curvature and torsion of the protein backbone by  representing it as a smooth line, which was defined by the set
of C$_\alpha$ atoms assumed to be equidistant. A 5th order spline was applied for the smoothing procedure but, contrary to the 3rd order spline used in APSA, this spline
does not pass through the chosen anchor points. Helices were described with high precision, but turns were not.    After subsequent normalization, the
curvature and torsion were used as signature parameters for structure comparison in different proteins. No secondary structure recognition was attempted.
Finally, it is noteworthy that  work on DNA helices has also been carried out utilizing a description with curvature and torsion. [39,40] 
\bigskip

\noindent {\bf 5. Conclusions and Outlook}
\smallskip

The description of the protein structure by its backbone is a well-accepted concept. The present work is based on this approach and
presents the backbone as a smooth line in 3D space rather than a collection of discrete backbone points. The natural
choice for the anchor points leading to a smooth backbone line are the C$_\alpha$ atoms because the latter represent the flexible
part of the backbone (N and C(=O) atoms are more constrained in their flexibility by the peptide bond), which directly reflects
on the conformational and steric influence of the side chains. By using a continuous rather than discrete representation of the
backbone we can project the 3D structure of the protein into the $\kappa(s)$ and $\tau(s)$ diagrams, which reveal
characteristic and easy to distinguish patterns for all secondary structural units of a protein. Protein structure analysis is
transferred thereby from a qualitative or semiquantative to a quantitative level: a) Different types of helices (3$_{10}$-,
$\alpha$-,  or $\pi$-helix), $\beta$-strands can be easily distinguished by $\kappa(s)$ and $\tau(s)$ diagrams. b) The regularity of
the structures can be measured and their distortions quantified. c) An analysis of amino acids occurring in ideal environments
reveals that only about 63 \% of DSSP's $\alpha$-helical amino acids and 48.5 \% in the case of $\beta$-strands are indisputably ideal. We suggest that the
disparity between various secondary structure assignment methods in the literature is partly due to this fact. d) The entire backbone can be
 presented as a combination of coiled and extended regions as shown for some turns.

APSA is based on the idea pf subsequent steps of coarse graining the protein backbone so that more and more nonlocal features are included into the description. In
subsequent work, [28,41] we have applied APSA to a whole range of tasks and problems occurring in connection with protein structure analysis. In Ref. 28a, APSA is used to
analyze 533
$\alpha$-helices and 644 $\beta$-strands taken from 77 proteins to show how natural secondary structures differ from ideal ones.
Kinks, distortions, and breaks are quantified and the boundaries (entry and exit) of secondary structures are classified. It is also shown how similar tertiary structures
can be easily recognized and similarity comparisons performed by analyzing the constituting helices, strands, turns, their entries and exits. In Ref. 28, additional
improvements are incorporated into the APSA protocol by extending the test set to 155 supersecondary structures. It is shown that the function $\tau$(s) is sufficient to
analyze the contributions of all amino acids to the conformation of the protein backbone. The characteristic peak and trough patterns of the $\tau$(s) diagrams can be
translated into a 16-letter code, which provides a rapid identification of helices, strands, and turns, specifies entry and exit points of secondary structural units, and
determines their regularity in terms of distortions, kinks or breaks. Via computer encoding, 3D protein structure is projected into a 1D string of conformational letters,
which lead to words (secondary structure), combination of words to phrases (supersecondary structure), and finally to whole sentences (representation of protein chains)
without loosing conformational details. The 3D-1D-projection procedure represents an extension of the APSA method.

A documentation of the list of structures described by APSA leads to a library of protein subunits
or building blocks that can be used in protein structure prediction and modeling. The analysis of the $\kappa$(s)and $\tau$(s) diagrams provides a meaningful and
systematic way of breaking down loop regions to substructures that contribute to the library of $\kappa$-$\tau$ patterns. Such libraries can be used to search an 
classify turns, supersecondary structures, and folds in a systematic way. [41]

The definition of folds and domains is often linked to
functionality, rather than to the conformation presented by the protein. Though this is aimed for far reaching goals, it leads to
classification systems that are currently semi-automated, producing conflicting results with the need of reviewing them in light of new folds, as is the case with SCOP
[7] and CATH [8]. APSA provides a unique way of solving this problem. At the tertiary level, $\kappa$(s) and $\tau$(s) rules can be used to recognize distinct folds,
define them, and set up databanks of coordinates for each. Another important application of APSA is similarity
comparison. APSA can handle rapid and accurate similarity measures at any user-defined level of coarseness: whether
an exact C$_\alpha$ to C$_\alpha$ match or determining localized equivalent regions between the two domains. Just as any line in 3D space can
be accurately and completely described by $\kappa$(s) and $\tau$(s), the availability of latter information could lead to the reconstruction of
the spline and ultimately, the backbone. In this way different segments of proteins can be put together into one continuous
backbone for further computational and folding studies. For example, such studies can be performed on special proteins that undergo
domain-swapping such as in the Alzheimers afflicted brain tissue. Other computational applications could include tracking, accurately and quantitatively, the
conformational changes in a protein as a result of ligand binding and protein-protein interaction.

\noindent {\bf Supporting Information available}: All curvature and
torsion diagrams, $\kappa(s)$ and $\tau(s)$, and the backbone line are given for each of the 20 proteins investigated in the Supporting Information.  

\bigskip
\vfill
\eject

\noindent {\bf  Acknowledgment}
\smallskip
\noindent EK and DC thank the National Science Foundation and the University of the Pacific for support of this work.

\bigskip
\bigskip

\noindent{\bf References}
\smallskip 

\item {[1]}  (a)  Pauling, J. L.; Corey, R.B; Branson, H.R. {\it Proc Natl Acad Sci, USA} {\bf 1951}, {\it 37}, 205-211.
	(b) Nelson, D. L.; Cox, M. M.;    {\it Lehninger Principles of Biochemistry, 5th Edition}, Freeman, New York, 2008.
	(c) Mathews, C.K.; van Holde, K.E.; Ahern, K.G. {\it Biochemistry}, Addison, Wesley Longman, New York, 1999.
 (d) Anderson, C. A. F.; Rost, B. in {\it Structural Bioinformatics}, Bourne, P. E.; Weissig, H., Edt.s, Wiley-Liss, Hoboken, New Jersey, 2003.

\item {[2]} Venkatachalam, C.M. {\it Biopolymers} {\bf 1968}, {\it 6}, 1425-36. 

\item {[3]} Kendrew, J.C.; Bodo, G.; Dintzis, H. M.; Parrish, R.G.; Wyckoff, H.; Phillipsa, D.C. {\it Nature}, {\bf 1958}, {\it 181}, 662-666. 

\item {[4]} Kabsch, W.; Sander, C. {\it Biopolymers}, {\bf 1983}, {\it 22}, 12, 2577-637. 

\item {[5]} Rao S.T.; Rossman M.G. {\it J. Mol. Bio.} {\bf 1973}, {\it 76}, 211-256.

\item {[6]} Westbrook, Z.; Feng, G.; Gilliland, T. N.; Bhat, H.; Weissig, I.N.; Shindyalov, P.E. {\it Nucleic Acids Research} {\bf 2000}, {\it 28}, 235-242. 

\item {[7]} Murzin, A.G.; Brenner, S.E.; Hubbard, T.; Chothia, C. {\it J. Mol. Biol.} {\bf 1995}, {\it 247}, 536-540. 

\item {[8]}  Orengo, C.A.; Michie, A.D.; Jones, S.; Jones, D.T.; Swindells, M.B.; Thornton, J.M. {\it  Structure} {\bf 1997}, {\it 5}, 1093-1108. 

\item {[9]} Dietmann, S.; Holm, L. {\it Nat. Struct. Biol.} {\bf 2001},{\it 8}, 953-957. 

\item {[10]} 	Day,R.; Beck, D.A.; Armen, R.S.; Daggett, V. {\it Protein Science} {\bf 2003}, {\it 12}, 2150-2160. 

\item {[11]} 	Frishman, D.; Argos, P. {\it Proteins}, {\bf 1995}, {\it 23}, 566-579.  

\item {[12]} 	Richards, F.M.; Kundrot, C.E. {\it Proteins}, {\bf 1988}, {\it 3}, 71-84.

\item {[13]} 	Martin, J.; Letellier, G.; Marin, A.; Taly, J.F.; Brevern, A.; Gibrat, J.F. {\it BMC Structural Biology}, {\bf 2005}, {\it 5}, 17.

\item {[14]}  Efimov, A.V.{\it  Prog. Biophys. Molec. Biol.} {\bf 1993}, {\it 60}, 201-239. 

\item {[15]} 	Richardson J., {\it Adv in Prot Chem.} {\bf 198}, {\it 34}, 167-339.

\item {[16]} 	Raveh, B.; Rahat, O.; Basri, R.; Schreiber, G. {\it Bioinformatics}, {\bf 2006}, {\it 23}, e163-e169. 

\item {[17]} 	Salamov, A.A.; Solovyev, V.V. {\it J. Mol. Biol.} {\bf 1995}, {\it 247}, 11-15. 

\item {[18]}	Kister, A.E.; Phillips, J.C. {\it Archiv.org, e-Print Archive, Condensed Matter} {\bf 2008}, 1-17.

\item {[19]} 	Bower, M.J.; Cohen, F.E.; Dunbrack, R.L. {\it J. Mol. Biol}, {\bf 1997}, {\it 267}, 1268-1282.

\item {[20]} 	Mattos, C.; Petsko, G.A.; Karplus, M. {\it J. Mol. Biol}, {\bf 1994}, {\it 238}, 733-747.

\item {[21]} Jones, D.T. {\it The Pharmacogenomics J.} {\bf 2001}, {\it 1}, 126-134. 

\item {[22]} Benedetti, E.; Blasio, B.D.; Pavone, V. Pedone, C.; Santini, A.; Crisma, M.; Toniolo, C. 
{\it Molecular Conformations and Biological Interactions}, Indian Academy of Sciences, Bangalore, 1991, p. 497-502.

\item {[23]} Emberly, E.G.; Mukhopadhyay, R.; Wingreen, N.S.; Tang, C. {\it J. Mol. Biol.}  {\bf 2003}, {\it 327}, 229-237.

\item {[24]} Kreyszig, E. {\it Differential Geometry, Dover Publications}, London, 1991.

\item {[25]}	Neumaier, A.; Dallwig, S.; Huyer, W.; Schichl, H. {\it New techniques for the construction of residue potentials for protein folding: 
Algorithms for Macromolecular Modeling. Lecture Notes Comput. Sci. Eng.} 4, Leimkuhler, B.; Chipot, C.; Elber, R.; Laaksonen, A.; Mark, A.; Schlick, T.; Sch\"utte, C.; 
Skeel, R., Eds., Springer, Berlin 1999, 212-224. 

\item {[26]}	Tung, C.-H.; Huang, J.-W.; Yang, J.-M. {\it Genome Biology} {\bf 2007}, {\it 8}, R31.

\item {[27]}	Arab S.; Didehvar, F.; Saleghi, M.; {\it Iranian J. Biotech.} {\bf 2007}, {\it 5}, 93-99.

\item {[28]}  (a) Ranganathan, S.; Izotov, D.; Kraka, E.; Cremer, D. {\it Proteins: Structure, Functions, and Genetics}, submitted.  arXiv:0811.3034v2 [q-bio.QM].
(b) Ranganathan, S.; Izotov, D.; Kraka, E.; Cremer, D.,  arXiv:0811.3258v2[q-bio.QM].

\item {[29]}	Konkoli, Z.;  Kraka, E.; Cremer, D. {\it J. Phys. Chem. A}  {\bf 1997}, {\it 101}, 1742.

\item {[30]}	Kraka, E. in {\it Encyclopedia of Computational Chemistry}, Schleyer, P.v.R.; Allinger, N.L.; Clark, T.; Gasteiger, J.; Kollman, P.A.; Schaefer III, H.F.; 
Schreiner, P.R. Eds. John Wiley, Chichester, UK, 1998, Vol. 4, p.2437.

\item {[31]}	Rackovsky, S.; Scheraga, H.A. {\it Macromolecules}, {\bf 1978}, {\it 11}, 1168-1174.

\item {[32]} 	Louie, A.H.; Somorjai, R.L. {\it J. Mol. Biol.} {\bf 1983}, {\it 168}, 13-162.

\item {[33]}	Louie, A.H.; Somorjai, R.L. {\it J. Theor. Biol.} {\bf 1982}, {\it 98}, 189-209.

\item {[34]} (a) 	Soumpasis, D.M.; Strahm, M.C. {\it J. Bimolecular Structure and Dynamics} {\bf 2000}, {\it 17}, 965-979. (b) Soumpasis, D.M.; Strahm, M.C.  {\it
Structure, Motion, Interaction and Expressions of Biological Macromolecules},  Proceedings of the Tenth Conversation, State University of New York, Albany, NY 1998,
11-32.

\item {[35]} Hausrath, A. C.; Goriely, A.; {\it J. Struct. Biol.} {\bf 2007}, {\it 158}, 267-281. 	

\item {[36]}	Sklenar, H.; Etchebest, C.; Laverey, R. {\it Proteins: Structure, Functions, and Genetics}, {\bf 1989}, {\it 6}, 46-60.

\item {[37]} 	Zhi, D.; Krishna, S.; Cao, H.; Pevzner, P.; Godznik, A. {\it BMC Bioinformatics}, {\bf 2006}, {\it 7}, 460.

\item {[38]}	Can, T.; Wang, Y.-F. {\it J. Bioinformatics and Comp. Biol.} {\bf 2004}, {\it 2}, 215-239.

\item {[39]}	Goyal, S.; Perkins, N.C.; Lee, C.L. {\it J. Comp. Phys.} {\bf 2005}, {\it 209}, 371-389.

\item {[40]}	Goyal, S.; Lillian, T.; Blumberg, S.; Meiners, J.-C.; Meyhofer, E.; Perkins, N.C. {\it Biophys. J.} {\bf 2007}, {\it 93}, 4342-4359.

\item {[41]}    Ranganathan, S.; Izotov, D.; Kraka, E.; Cremer, D.,  arXiv:0811.3464v2 [q-bio.QM]

\bigskip
\bigskip

\vfill \eject

\noindent {\bf APPENDIX I}
\smallskip

 For the purpose of determining the number of amino acid residues
influenced by the natural boundary conditions, a test was performed on 1HVZ, a circular polypeptide taken from the PDB [6]. This
cyclic protein with 18 amino acids (residue 18 is connected to residue 1; see Figure A1) was analyzed  treating residue 1 (and 8) both as a terminal
and as a central residue. For this purpose the analysis was started as residue 1 and  performed through residues 2, $\cdots$, 8, $\cdots$, 18, 1, $\cdots$, and stopped at
residue 8 being now a terminal residue, i.e. residue 1 (8) functioned both as a terminal and central residue.

  A comparison of the curvature of the repeated regions  (Figures A1a) and A1b)) reveals the error due to boundary conditions. Curvature and torsion for the first two
residues differ significantly whereas the error for the third residue is relatively small. THis is confirmed for the backbone line of 1HVZ shown in Figure A1c. The spline
lines of the two passes in regions G1 to R8 coincide with the exception of the parts  G1-F2 (start of the analysis) and C7-R8 (termination of the analysis). 
It was inferred that the spline is reliably described from the third amino acid onwards whereas errors for the  residue next to the terminal one are moderate, however
substantial for the terminal residue itself. We will exclude therefore terminal and the two next-to-terminal residues  of a protein from the analysis.
\bigskip


\centerline{\epsfbox {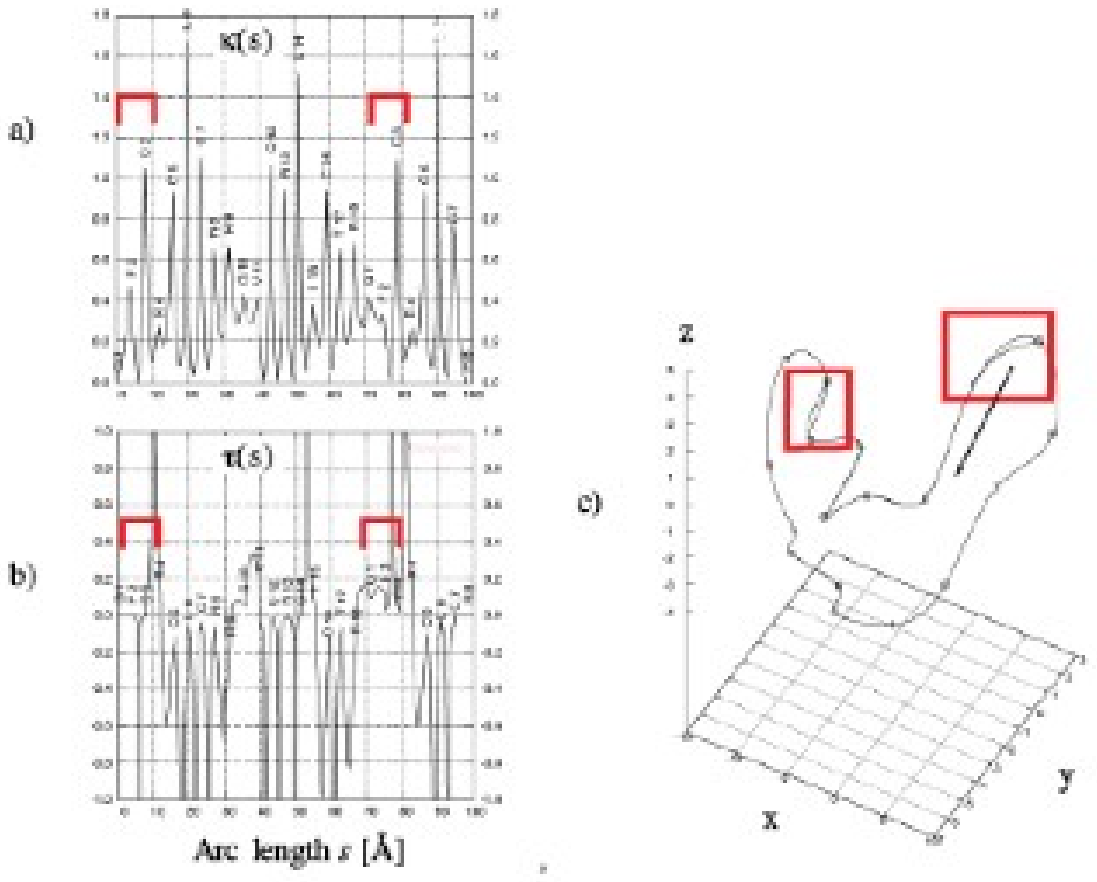}}

\medskip

\noindent {\bf Figure A1:} Curvature and torsion analysis for the cyclic polypeptide 1HVZ. a) Curvature diagram, b) torsion diagram, and c) backbone line for a pass from
residue 1  thru residues 2, $\cdots$, 8, $\cdots$, 18, 1, $\cdots$, 8 (see text). The region of deviations due to the boundary conditions are given in red.

\bigskip
\bigskip

\bye

\noindent {\bf APPENDIX II}
\smallskip

  The theory of spline interpolation is well established (see e.g. Ref.~[24]). In the present paper, the cubic
spline interpolation with the natural boundary conditions was employed for curvature and torsion calculations. Since
the spline  passes through the anchor points (knots), the curvature and torsion computed from it might be prone to
variations in the  coordinates of the anchor points. In reality the coordinates of the knot atoms are always determined with some
uncertainty. Because of this, it is useful to determine how sensitive  curvature and torsion of the spline-fitted backbone of a
protein are to small variations in the knot coordinates.

\midinsert

\centerline{\epsfbox {Appendix2.eps}}
 
\medskip

\noindent {\bf Figure A2:} Variation of the spline curvature with the change in the y-coordinate of the knot, $y_i
\rightarrow y_i +\epsilon$, $\epsilon = (a) 10^{-4}, (b)  10^{-3}, and (c) 10^{-2}$, at $x \approx 0.8$ for $y(x) = \sin(x)$. The
number of knots is $n = 20$ for $h = 0.157$.
\endinsert

For illustration, the planar curve, $y(x) = \sin x$, $x \in [0,\pi]$, is chosen. It exhibits typical curvature features
of a protein segment and, apart from this, it is easy to analyze. The number of knots is set to $n = 21$ and the distance
between the knot points is given by $h = \pi / n \approx 0.157$. Figure~A1 shows the curvature variation when changing the
y-coordinate of a single knot point at $x \approx 0.8$ according to $y \rightarrow y + \epsilon$. The deviation $\epsilon$ was
set to $0.0001$, $0.001$, and $0.01$, respectively. As one can see, the variation in curvature grows  quickly with $\epsilon$.

It would be incorrect to apply the $\epsilon$ values used in the above example directly to the uncertainties in the
atomic coordinates because $\delta \kappa$ also depends on the distance $h$ between the knot points. For a given
$\epsilon$, $\delta \kappa \rightarrow \infty$, when $h \rightarrow 0$. For the purpose of making a reliable estimate of the tolerable
uncertainty in the  coordinates of the anchor points, one has to find a relationship between $\delta \kappa$, $\epsilon$, and $h$.
For this purpose,  a planar cubic piece-wise spline $y(x) = \bigcup_{i=1}^{n-1} y_i(x)$ is considered where 
$$ 
y_i(x) = a_i + b_i (x-x_i) + c_i (x-x_i)^2 + d_i (x-x_i)^3, \quad i=1,\dots,n-1,
\eqno(A1)
$$ 
with $a_i$, $b_i$, $c_i$, $d_i$ being the spline coefficients, $x \in [X_1,X_2]$  the parameter, and
$\{x_i\}_{i=1}^n$, $x_1=X_1$, $x_n=X_2$ some partitioning of the interval $[X_1,X_2]$.  Usually, the partitioning
grid is uniform, $x_{i+1} - x_i = h$, $i=0,\dots,n-1$. The spline coefficients can be found from the requirement
that the spline passes through the knot points. 
The curvature of a planar curve is defined by
$$
\kappa(x) = {|y''| \over (1+y'^2)^{3/2}} \eqno(A2).
$$ 
Using these definitions, one finds that $\delta \kappa \propto \epsilon / h^\alpha$,
$\alpha = 2$ for small $h$. Figure~A2 shows the dependence $\delta\kappa$ as a function of $1/h$. From the graph one
can confirm that $\alpha \approx 2$ and $\delta \kappa \approx 2\epsilon / h^2$.

\medskip
\centerline{\epsfbox {Appendix3.eps}}

\smallskip

\noindent {\bf Figure A3:} Uncertainty in the curvature, $\delta \kappa$ in dependence of  the number {\it n} of anchor points (knot points), i.e.
the distance {\it h} between the knot points (for increasing {\it n}, distance {\it h} decreases and $\delta \kappa$ becomes large).
\medskip
If one chooses the C$_\alpha$ atoms as the spline knots, then $h \approx 4$~\AA{}. For quantitative analysis
$\delta\kappa$ must be less than $0.1$~\AA$^{-1}$. This gives $\epsilon_{\rm max} \approx 1$~\AA{} as an upper limit
for coordinates uncertainty that can be tolerated. If the coordinate uncertainties $\epsilon < \epsilon_{\rm max}$,
the uncertainty in curvature $\delta\kappa < 0.1$~\AA.

\vfill \eject

\bye
\medskip
\centerline{\TableA}
\medskip

\bigskip
\bigskip

\medskip
\centerline{\TableB}
\medskip

\vfill \eject

\noindent Figure 1

\medskip
\centerline{\FigureA} 
\medskip

\bigskip
\vskip 2 cm

\noindent Figure 2

\medskip
\centerline{\FigureB} 
\medskip

\bigskip
\vfill \eject

\noindent Figure 3

\centerline{\FigureCA}
\centerline{\FigureCB}
\centerline{\FigureCC}
\medskip

\bigskip
\bigskip

\noindent Figure 4

\medskip
\centerline{\FigureD} 
\medskip

\vfill \eject

\noindent Figure 5

\medskip
\centerline{\FigureE} 
\medskip

\vfill \eject

\noindent Figure 6

\medskip
\centerline{\FigureF} 
\medskip

\vfill \eject

\noindent Figure 7

\medskip
\centerline{\FigureG}
\medskip

\bye